\documentclass[]{tandf2}

\usepackage{epstopdf}
\usepackage{subfigure}

\usepackage[numbers,sort&compress]{natbib}
\bibpunct[, ]{[}{]}{,}{n}{,}{,}
\renewcommand\bibfont{\fontsize{10}{12}\selectfont}
\makeatletter
\def\NAT@def@citea{\def\@citea{\NAT@separator}}
\makeatother

\theoremstyle{plain}

\theoremstyle{definition}

\theoremstyle{remark}

\begin{document}

\articletype{ARTICLE TEMPLATE}

\title{Chemical\--Diffusive Models for Flame Acceleration andÊ
Transition to Detonation: Genetic Algorithm and Optimization Procedure}


\author{
\name{C.R.\ Kaplan$^{\ast}$\thanks{$^\ast$Corresponding author. Email: crkaplan@umd.edu}, A.\ \"{O}zgen and E.S.\ Oran}
\affil{Department of Aerospace Engineering, University of Maryland, College Park, MD}
}

\maketitle

\begin{abstract}
One of the most important and difficult parts of constructing a multidimensional numerical simulation of flame acceleration and deflagration-to-detonation transition (DDT) in a reacting flow is finding a reliable and affordable model of the chemical and diffusive properties. For simulations of realistic scenarios, full detailed chemical models (with hundreds of chemical reactions and many species) are computationally prohibitive.  In addition, they are usually inaccurate for high-temperature and high-pressure shock-laden flows. This paper presents a general approach for developing an automated procedure to determine the reaction parameters for a simplified chemical-diffusive model to simulate flame acceleration and DDT in stoichiometric methane-air and ethylene-oxygen mixtures.  The procedure uses a combination of a genetic algorithm and Nelder-Mead optimization scheme to find the optimal reaction parameters for a reaction rate based on an Arrhenius form for conversion of reactants to products.  The model finds six optimal reaction parameters (ratio of specific heats, activation energy, preexponential factor, heat release rate, thermal conductivity coefficient, and overall molecular weight) that reproduce six target flame and detonation properties (adiabatic flame temperature, constant volume equilibrium temperature, laminar flame speed, laminar flame thickness, Chapman-Jouguet detonation velocity, and detonation half-reaction thickness).  Results from the optimization procedure show that the optimal reaction parameters, when used in 1-D reactive Navier-Stokes simulations, closely reproduce the target flame and detonation properties for the stoichiometric methane-air and ethylene-oxygen mixtures.  The effects of uncertainties in the values of target flame and detonation properties can be minimized to have little effect on the resulting optimal reaction parameters, and the reaction parameters can be tailored, if necessary, for the different regimes of flame acceleration and DDT.  When the reaction parameters are used as input in a 2-D simulation of flame acceleration and DDT in an obstacle-laden channel containing stoichiometric methane-air, the simulation results closely follow the transition to detonation observed in experiments.   This automated procedure for finding parameters for a proposed reaction model makes it possible to simulate the behavior of flames and detonations in large, complex scenarios, which would otherwise be an incalculable problem.
\end{abstract}


\section{Introduction}
\label{intro}

Deflagration-to-detonation transition (DDT) appears as a sudden change in the propagation mode of a combustion wave from an accelerating deflagration  to a detonation~\cite{oran_origins_2007}. This transition usually occurs in the presence of shock waves, turbulence, and boundary layers. After the transition to a detonation, there are substantial increases in the velocity of the combustion wave and in the temperatures and pressures that follow it. For this reason, DDT is both dangerous if it occurs accidentally in industrial environments, such as coal mines and fuel storage facilities \cite{zipf_2013,zipf_report,ciccarelli_2008}, and  useful  when the energy it produces can be harnessed for propulsion devices, such as pulse detonation engines~\cite{roy_pulse_2004}, microscale thrust generators~\cite{ju_microscale_2015}, and rotating wave detonation engines \cite{nordeen_rde}. In the context of astrophysical combustion, DDT can help explain the origin of supernovas \cite{oran_origins_2007}.
			
There have been many computational studies performed to clarify the physical mechanisms leading to flame acceleration and DDT. These studies require appropriate numerical models of the combination of fluid dynamics, chemical reactions and energy release, and a number of physical diffusion processes, such as molecular diffusion and thermal conductivity. The basis for the solution is the set of unsteady compressible Navier-Stokes equations, which need to be solved ``accurately enough'' to resolve acoustic, shock, and turbulence phenomena and all of their interactions. 
			
One major problem in solving these equations is finding adequate models of the chemical reactions, heat release, and physical diffusion processes that are required source terms for modeling flames and detonations using the Navier-Stokes equations. For many  gaseous fuels, the most complex and detailed chemical reaction mechanisms proposed have been sets of elementary reactions that attempt to describe all of the major and intermediate chemical species and all of the chemical interactions among them. Such reaction mechanisms can become inordinately expensive to use when included in a simulation of DDT.
For example, the mechanism proposed by Wang et al. \cite{wang_comprehensive_1998} for ethylene and acetylene combustion contains 50 reacting species and over 350 chemical reactions, and that proposed for sooting ethylene flames by Kazakov et al. \cite{kazakov_1995} contains 100 species and 500 reactions. Calculations including this many reaction rates and tracking changes in concentration of this many reacting species increase the CPU requirement dramatically for adequately resolved unsteady shock-laden multidimensional calculations. This has led to the question of whether simpler chemical-diffusive models suffice.
			
But there is another issue that is perhaps more challenging, and one that cannot be addressed as easily as saying, ``Well, we just need  a larger computer.'' The currently available chemical reaction mechanisms, to say nothing of diffusive processes, are inadequate for describing the high temperatures and pressures encountered in shock-laden, high-temperature, high-pressure environments that arise before and through DDT and in a propagating detonation. As shown by Taylor et al. \cite{taylor_numerical_2013}, these conditions lead to situations in which excited molecular states are formed, and interactions among molecules in excited states may occur. These types of processes may alter the structure and timing of energy-release during the autoignition period, a topic that requires considerably more examination and research. 

In summary, it appears that currently available detailed chemical reaction mechanisms are both computationally prohibitive and inaccurate for high-temperature and high-pressure shock-laden flows.   Thus, there is now an important question to address: {\sl What type and level of chemical-diffusive models would be adequate to use for modeling flame acceleration and DDT? }  In prior work, we made a significant effort to develop what we believe is an adequate level of representation of the chemical-diffusive processes for computing the initial flame development through DDT. This approach involves a simplified exponential form that looks like an Arrhenius reaction with a series of adjustable, calibrated parameters. The generic form of this reaction is:

\begin{equation}
{\dot \omega}  =  - A\rho Y{{\mathop{\rm e}\nolimits} ^{ - E_a/RT}}
\label{arrhenius}
\end{equation} 
\noindent
where $\dot \omega$ is the reaction rate, $A$ is the pre-exponential factor, $\rho$ is the fluid density, $Y$ is the fuel mass fraction, $E_a$ is the activation energy, $R$ is the universal gas constant, and $T$ is the fluid temperature.  In addition, it is necessary to define physical diffusion parameters to model laminar flame propagation.  This approach was used for flame acceleration and DDT by Khokhlov and Oran \cite{khokhlov_numerical_1999}, where the model parameters were set for  low-pressure acetylene-air, by Gamezo et al. \cite{gamezo_influence_2001} for low-pressure ethylene-air mixtures, by  Gamezo et al. \cite{gamezo_2007,gamezo_2008,gamezo_deflagration--detonation_2009} and Ogawa et al. \cite{ogawa_2013a,ogawa_2013b} for hydrogen-air mixtures at atmospheric conditions, and recently by Kessler et al. \cite{kessler_simulations_2010,kessler_2012} for atmospheric methane-air. The sum of these computations and the surprisingly good comparison with results from large- and small-scale experiments has given us confidence that this is, in fact, a valid approach for large-scale computations involving complex physical processes and their interactions.

When this approach was originally used, the only attempt was to use ``physically reasonable'' model parameters. The results were expected to be qualitatively interesting, but not quantitative. Subsequently, when it was shown that the approach was giving quantitatively correct descriptions of DDT, efforts were made to optimize the parameters to select the best representations of flames and detonations, as determined from experiments, theory, and with input from existing detailed chemical reaction models. For example,  Gamezo \cite{gamezo_deflagration--detonation_2009} and Kessler \cite{kessler_simulations_2010} discuss approaches to finding good model input parameters, when, in fact, there are many sets that reproduce the correct overall flame and detonation properties.

This paper presents a general approach for developing an automated procedure to determine the input parameters for a model chemical-diffusive system. The procedure is based on using a standard form of conversion from reactants to products, as given in equation \ref{arrhenius}, and using a combination of a genetic algorithm and an optimization scheme to find the best parameters \cite{alp_scholarly_paper}. Here we describe the procedure in some detail and then use it to find parameters for stoichiometric methane-air and ethylene-oxygen mixtures at atmospheric conditions.  Use of the procedure to find reaction parameters for other fuels, other mixtures with spatially varying equivalence ratios, and more complex (possibly multistep) forms for representing the evolution of the energy release, will be described in future papers.

\section{Background}

\subsection{Reactive-Flow Navier-Stokes Equations}
\label{NSE}

Simulation of flame acceleration and DDT in obstacle-laden channels containing fuel-oxidizer mixtures is carried out by solving the unsteady, fully compressible, reactive Navier-Stokes equations:

\begin{equation}
\frac{{\partial \rho }}{{\partial t}} + \nabla  \cdot (\rho {\bf{U}}) = 0 ,
\end{equation}
\begin{equation}
\frac{{\partial (\rho {\bf{U}})}}{{\partial t}} + \nabla  \cdot (\rho {\bf{UU}}) + \nabla P + \nabla  \cdot {\bf{\hat \tau }} = 0 ,
\end{equation}
\begin{equation}
\frac{{\partial E}}{{\partial t}} + \nabla  \cdot ((E + P){\bf{U}}) + \nabla  \cdot ({\bf{U}} \cdot {\bf{\hat \tau }}) + \nabla  \cdot (K\nabla T) + \rho q\dot \omega  = 0 ,
\end{equation}
\begin{equation}
\frac{{\partial (\rho Y)}}{{\partial t}} + \nabla  \cdot (\rho Y{\bf{U}}) + \nabla  \cdot (\rho D\nabla Y) - \rho \dot \omega  = 0 ,
\end{equation}
\begin{equation}
{\bf{\hat \tau }} = \rho \nu \left( {\frac{2}{3}(\nabla  \cdot {\bf{U}}){\bf{I}} - (\nabla {\bf{U}}) - {{(\nabla {\bf{U}})}^\dag }} \right) ,
\end{equation}

\noindent closed with the ideal gas equation of state

\begin{equation}
\rho  = \frac{{P{M_w}}}{{RT}}.
\end{equation}

\noindent In these equations, $t$ is time, $\bf{U}$ is the velocity vector, $P$ is pressure, $\tau$ is the viscous stress tensor, $E$ is the fluid energy density,  $K$ is the thermal conductivity, $T$ is temperature, $q$ is the chemical energy release, $Y$ is the mass fraction of reactants, $D$ is the diffusion coefficient, $\nu$ is viscosity, $\bf{I}$ is the identity matrix, $M_w$ is the molecular weight, and the superscript $\dag$ is the transpose notation for a matrix.  

As discussed in Section \ref{intro}, the level of representation of the chemical reaction terms, the $\dot\omega$, can vary from detailed chemical reaction mechanisms to simplified forms.  In this work, we extend the approach used in previous studies \cite{khokhlov_numerical_1999,gamezo_influence_2001,gamezo_2007,gamezo_2008,gamezo_deflagration--detonation_2009, ogawa_2013a,ogawa_2013b,kessler_simulations_2010,kessler_2012}, in which the reaction rate is based on a single equation with an Arrhenius form representing the conversion of a mixture of fuel and oxidizer to product: 

\begin{equation}
\frac{{dY}}{{dt}} = \dot \omega  =  - A\rho Y{{\mathop{\rm e}\nolimits} ^{ - E_a/RT}}  .
\end{equation}

\indent During flame acceleration leading to DDT, the temperature changes substantially, so viscosity ($\nu$), mass diffusivity ($D$), and thermal diffusivity ($\alpha$) have the following temperature dependent form:

\begin{equation}
\nu  = {\nu _0}\frac{{{T^n}}}{\rho },
\label{visco}
\end{equation}
\begin{equation}
D = {D_0}\frac{{{T^n}}}{\rho },
\label{diffu}
\end{equation}
\begin{equation}
\alpha = \frac{K}{{\rho {c_p}}} = {\kappa _0}\frac{{{T^n}}}{\rho }.
\label{thermal_diffu}
\end{equation}

\noindent In these expressions, $\nu _0$, $D_0$, and $\kappa _0$ are constants, and $c_p$ is the heat capacity at constant pressure, 

\begin{equation}
c_p = \frac{\gamma R}{{M_w}(\gamma-1)}
\label{eq_cp}
\end{equation}

\noindent
where $\gamma$ is the specific heat ratio.  The exponent $n$ is chosen to be 0.7, which has been shown to emulate  the typical temperature dependence for the system  \cite{kessler_simulations_2010}.

The set of equations given above require certain input parameters, namely $\gamma$, $A$, $E_a$, $q$, $\kappa_0$, $D_0$, $\nu_0$, and $M_w$. These parameters define the chemical and diffusive properties of the gas and control the amount and rate of heat release locally and temporally within the flow.  By setting the Lewis number equal to unity and using a user-defined value of the Prandtl number, the list of required input reaction parameters reduces to six variables: $\gamma$, $A$, $E_a$, $q$, $\kappa_0$, and $M_w$.  These are the six reaction parameters used in this study.

As will be discussed in detail in Section \ref{optimization_procedure} below, the optimization procedure ensures that when these six reaction parameters are used to find solutions to the full set of Navier-Stokes equations, they should reproduce important properties of the flame and detonation.   These flame properties include the adiabatic flame temperature ($T_b$), laminar flame speed ($S_l$), laminar flame thickness ($x_{ft}$),  and constant-volume equilibrium temperature ($T_{cv}$), while the detonation properties include the Chapman-Jouguet detonation velocity ($D_{CJ}$), and half reaction thickness ($x_d$).

In the following section, we discuss a graphical approach which has been used \cite{kessler_simulations_2010} to find a set of reaction parameters for a stoichiometric methane-air mixture.  Then, in Section \ref{optimization_procedure}, we present a new approach that uses a combination of genetic and optimization algorithms to find optimal reaction parameters for stoichiometric mixtures of both methane-air and ethylene-oxygen.  

\subsection{Graphical Approach for Determination of Reaction Parameters}
\label{kessler_method}
\begin{figure}[!htbp]
		\centering
		\includegraphics[width=0.8\textwidth]{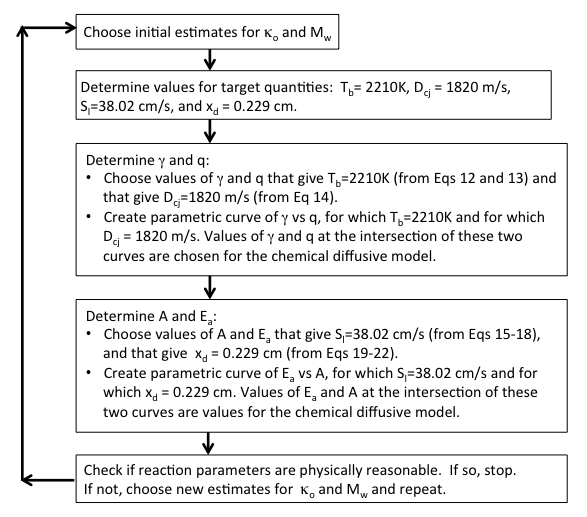}
		\centering
		\caption{Graphical approach procedure from Ref \cite{kessler_simulations_2010}}
		\label{kessler_graph_approach}
\end{figure}

Here we use the term ``graphical approach'' to describe the procedure developed by Kessler et al. \cite{kessler_simulations_2010} to find the reaction parameters for a stoichiometric methane-air mixture.  The model parameters are found by creating curves of constant properties of the mixture, and then choosing parameters at the intersection of the curves.  This procedure searches for a set of reaction parameters, $\gamma$, $A$, $E_a$, $q$, $\kappa_o$, and $M_w$ for which the computed values of $T_b$, $S_l$, $D_{CJ}$, and $x_d$ match a set of prespecified target values.  This procedure is represented in Fig. \ref{kessler_graph_approach}  and described below.  {\sl{The graphical approach \cite{kessler_simulations_2010} is discussed here in detail because it gives the background for the more complex approach that follows in section \ref{optimization_procedure}.}} 

First, initial values of $\kappa_o$ and $M_w$ are estimated.  Then target values of $T_b$, $S_l$, $D_{CJ}$ and $x_d$ are selected.  For example, for stoichiometric methane-air mixtures, $T_b$ = 2210K, $S_l$ = 38.02 cm/s, $D_{CJ}$ = 1820 m/s, and $x_d$ = 0.229 cm are reasonable values based on experimental data and computations using existing detailed chemical mechanisms.   

Next, values for $q$ and $\gamma$ are determined, as they are related to $T_b$ for laminar flames \cite{glassman_2008} and to $D_{CJ}$ for detonations \cite{zeldovich_1950}, by: 

\begin{equation}
{T_b} = {T_0} + \frac{q}{{{c_p}}}.
\label{eq:tb_heatrelease}
\end{equation}

\begin{equation} 
{D_{CJ}} = {c_0}\left( {\sqrt {1 + \frac{q}{{{P_0}}}\frac{{{\rho _0}({\gamma ^2} - 1)}}{{2\gamma }}}  + \sqrt {\frac{q}{{{P_0}}}\frac{{{\rho _0}({\gamma ^2} - 1)}}{{2\gamma }}} } \right)
\label{eq_dcj}
\end{equation}

\noindent
where $c$ is the sound speed, and the subscript {\sl{o}} corresponds to  ambient conditions.  As indicated by the third block in Fig.\ref{kessler_graph_approach}, various combinations of $q$ and $\gamma$ that give the target $T_b$=2210K are calculated from Eqs.~\ref{eq_cp} and \ref {eq:tb_heatrelease}, and combinations of $q$ and $\gamma$ that give the target $D_{CJ}$=1820 m/s are calculated from Eq.~\ref{eq_dcj}.  Then, on a graph of $\gamma$ vs. $q$, two parametric curves are drawn:  one curve represents values of $q$ and  $\gamma$ for which $T_b$ =2210K and the other curve represents values of $q$ and  $\gamma$ for which $D_{CJ}$=1820 m/s.   The values of $q$  and $\gamma$ at the intersection of these two curves are the values chosen for the chemical-diffusive model.  As indicated in Ref \cite{kessler_simulations_2010}, these values are $\gamma$=1.197 and  $q$=39.0${RT_o}/{M_w}$. 

Once $q$ and $\gamma$ are determined, a similar procedure is used to determine $A$ and $E_a$, as shown in the fourth block in Fig.\ref{kessler_graph_approach}.  The properties of a 1-D laminar flame are computed by solving Eqs.~\ref{flame1}-\ref{flame3}, which describe thermal conduction and energy release inside a steady state reaction wave.  An iterative procedure, discussed in detail in Ref \cite{kessler_simulations_2010}, is used to determine $S_l$ such that the computed 1-D flame profile satisfies the condition that $T=T_b=2210K$ when the 1-D temperature gradient becomes zero. This procedure is used to determine the values of $E_a$ and $A$ for which $S_l$ = 38.02 cm/s.   

\begin{equation}
\frac{{d{F_t}}}{{dx}} = \rho \left( {{U_l}{c_p}\frac{{dT}}{{dx}} - q\dot \omega } \right),
\label{flame1}
\end{equation}

\begin{equation}
{F_t} = K\frac{{dT}}{{dx}},
\label{flame2}
\end{equation}

\begin{equation}
K = {\kappa _0}{T^{0.7}}{c_p}.
\label{flame3}
\end{equation}

\indent In these equations, $F_t$ is the conductive heat-transfer flux, and $U_l$ is the velocity of the flame at a given point along $x$. Since the flame is assumed to be planar, $U_l$ can be found from the continuity equation,

\begin{equation}
{U_l} = \frac{{{S_l}{\rho _0}}}{\rho },
\end{equation}

Finally, the properties of a 1-D detonation are computed using a Zeldovich-von Neumann-Doering (ZND) model \cite{zeldovich_1950, vonneumann_1963, doring_1943}.  Values for $P_{ZND}$, $\rho_{ZND}$ and $e_{ZND}$ are obtained from expressions for a 1-D planar shock wave moving at $D_{CJ}$.  These ZND parameters are used as initial conditions for the integration of equations describing the reaction zone of a 1-D detonation, Eqs.~\ref{zndstart}-\ref{zndend}, from the initial conditions to the position where the detonation velocity reaches the sound speed.  

\begin{equation}
\frac{{d\rho }}{{dt}} = \frac{{q\dot \omega \rho (\gamma  - 1)}}{{{U^2} - {c^2}}},
\label{zndstart}
\end{equation}

\begin{equation}
\frac{{dE}}{{dt}} = \frac{P}{{{\rho ^2}}}\frac{{d\rho }}{{dt}} + q\dot \omega, 
\end{equation}

\begin{equation}
\frac{{dx}}{{dt}} = U,
\end{equation}

\begin{equation}
U = \frac{{{D_{CJ}}{\rho _0}}}{\rho },
\label{zndend}
\end{equation}

\noindent
The ZND reaction zone profile is then used to find the half-reaction thickness, $x_d$, defined as the distance between the leading shock wave and the point where half of the fuel has been consumed in the flame zone.  This procedure is used to determine the values of $A$ and $E_a$ for which $x_d$ =0.229 cm.  On a graph of $E_a$ vs. $A$, two curves are drawn:  one showing values of $E_a$ and $A$ for which $x_d$=0.229 cm and another showing values of  $E_a$ vs. $A$ for which $S_l$ =38.02 cm/s.  The values of $A$ and $E_a$ at the intersection of these two curves are the values chosen for the chemical diffusive model.  As indicated in Ref \cite{kessler_simulations_2010}, these values are $A$=1.64x${10^{13}}$cm$^3$/g$\cdot$s and $E_a$=67.55${RT_o}$.


Once values for $A$, $E_a$, $q$ and $\gamma$ are determined, it is necessary to check these to ensure that these reaction parameters are physically reasonable.  Examples of possible unphysical parameters include instances where the activation energy may be negative, or if the computed $\gamma>1.4$ or $\gamma<1.0$.   It is possible for the reaction parameters to be unphysical if the initial estimates of $\kappa_o$ or $M_w$ are inaccurate.  Another issue that may lead to unphysical reaction parameters is due to the fact that the method determines six reaction parameters ($\gamma$, $A$, $E_a$, $q$,  $\kappa_o$ and $M_w$) by matching four target values ($T_b$, $S_l$, $D_{CJ}$, and $x_d$).  This inequality can lead to nonunique solutions, such that more than one set of reaction parameters can produce the given target values.  As indicated in the last block of Fig. \ref{kessler_graph_approach}, if any of the reaction parameters are physically unreasonable, then new values for $M_w$ or $\kappa_o$ are chosen, or the target values (for $T_b$, $S_l$, $D_{CJ}$, and $x_d$) are changed, and the calibration procedure is repeated.

\section{Optimization Procedure}
\label{optimization_procedure}

As discussed above, the graphical approach can give unphysical values for the model parameters.  {\sl{To avoid this problem and make it possible to generalize the model equations, we have developed an automated procedure, based on a combination of a Genetic Algorithm and Nelder-Mead optimization scheme.}}  This new procedure, called GA-NM, is presented below.  

As before, we need to determine values for the six reaction parameters $\gamma$, $A$, $E_a$, $q$, $\kappa_o$ and $M_w$ that reproduce prespecified flame and detonation properties of the mixture.  Now, we have included two additional target properties, so that there are six target quantities and six reaction parameters. That is, in addition to the target quantities $T_b$, $S_l$, $D_{CJ}$, and $x_d$ matched in the graphical method, we also include the laminar flame thickness, $x_{ft}$, and constant volume equilibrium temperature, $T_{cv}$, defined as:

\begin{equation}
{x_{ft}} = \frac{{{T_b} - T}}{{\max \left( {\frac{{dT}}{{dx}}} \right)}}.
\label{xf}
\end{equation}

\begin{equation}
{T_{cv}} = {T_0} + \frac{q}{{{c_v}}},
\label{Tcv}
\end{equation}

\noindent
where $c_v$ is heat capacity at constant volume, ${c_v} = {R}/{{{M_w}(\gamma  - 1)}}$.  $T_{cv}$ is chosen because DDT is generally a result of autoignition of the mixture, and autoignition is often modeled as a constant-volume process~\cite{oran_origins_2007,khokhlov_numerical_1999}. This way, the formulation takes into account  the effects of constant volume combustion.

\subsection{Application of a Genetic Algorithm}

A genetic algorithm \cite{goldberg_genetic_1989} adapts principles of Darwinian natural selection to find an optimum or viable state.  The theory is based on a principle of ``survival of the fittest,'' where individuals in a population with the most appropriate or optimal traits have a higher chance of survival and of breeding to produce new individuals.  In the context of development of this chemical-diffusive model, each ``individual'' in a population has a set of traits corresponding to the reaction parameters $\gamma$, $A$, $E_a$, $q$, $\kappa_o$ and $M_w$. As the genetic algorithm proceeds, some of these traits are passed along to future ``generations,'' some are completely discarded, and others mutate, such that an optimal set of traits is determined.

\begin{figure}[!htbp]
		\centering
		\includegraphics[width=1.0\textwidth]{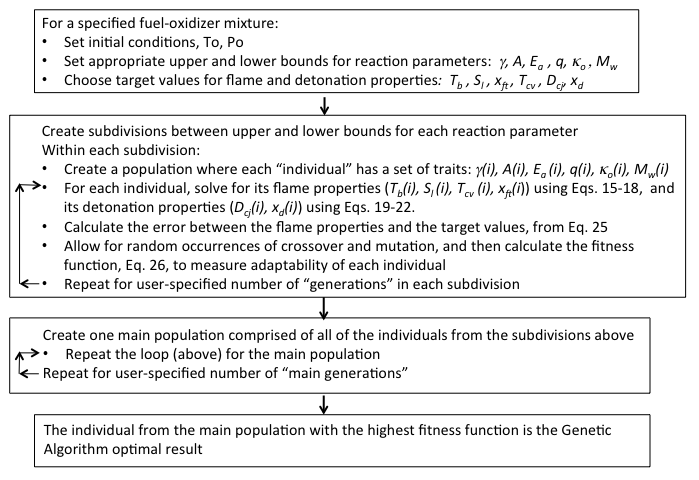}
		\centering
		\caption{Overall procedure for the Genetic Algorithm}
		\label{overall_ga}
\end{figure}

The new procedure for finding parameters for the chemical diffusive model is shown in Fig. \ref{overall_ga}.   For a specific fuel and oxidizer mixture,  initial ambient conditions (e.g., $T_o=298K$ and $P_o= 1 atm$) are selected, and prespecified target flame ($T_b$, $S_l$, $x_{ft}$, and $T_{cv}$) and detonation ($D_{CJ}$, $x_d$) properties were calculated using the Gordon McBride chemical equilibrium software \cite{Gordon_McBride}, Cantera \cite{Cantera}, and the Shock and Detonation Toolbox \cite{california_institute_of_technology_shock_2015}.  

As indicated in the second block of Fig.\ref{overall_ga}, a population is created in which each individual has a set of traits, $\gamma$, $A$, $E_a$, $q$, $\kappa_o$, and $M_w$.  
For each individual, the value of each trait is chosen randomly (using normally distributed random numbers) within an upper and lower bound.  Because the upper and lower bounds cover a large range, each range for each trait is also subdivided, to ensure that there are enough individuals covering the full range of values.  For the methane-air and ethylene-oxygen calculations discussed here, the upper and lower bounds and number of subdivisions for each parameter are shown in Table \ref{table_bounds}.

\begin{table}
 \tbl{Upper and lower bound for each reaction parameter, and the number of subdivisions for each.}
{\begin{tabular}{c c} \toprule 
Upper and Lower Bound & Number of Subdivisions \\
for Each Trait &  \\
  \midrule
  1.17 $<\gamma<$ 1.30 &  2 \\
   20 $<E_a/{RT_o}<$ 100 & 4 \\
30 $<qM_w/{RT_o}<$80 & 2 \\ 
1.0 $\times$ ${10^{10}}$ $<A<$ 1.0x${10^{16}}$  & 5 \\
1.0 $\times$ ${10^{-6}}$ $<\kappa_o<$ 1.0x${10^{-5}}$  & 2 \\
23 $<M_w<$ 30 & 2 \\
    \bottomrule
    \end{tabular}}
\label{table_bounds}
\end{table}

Then, separate genetic-algorithm calculations are carried out within each subdivision.  For each individual in each subdivision, using the individual's values of $\gamma$, $A$, $E_a$, $q$, $\kappa_o$, and $M_w$, the flame properties are found by solving Eqs.\ref{flame1}-\ref{flame3}, and the detonation properties are found by solving Eqs.\ref{zndstart}-\ref{zndend}.  These calculated flame and detonation properties are then compared to the prespecified target flame and detonation properties (calculated from chemical equilibrium software), and an error is defined as:

\begin{equation}
\label{error}
{Error}=\left[ \sum_{i=1}^{ntargets} \left(\frac {\xi_i-\xi_{i,target}} {\xi_{i,target}} \right)^2 \right]^{1/2}
\end{equation}

\noindent 
where $\xi_i$ represents each trait ($T_b$, $S_l$, $x_{ft}$, $T_{cv}$, $D_{CJ}$, and $x_d$).
A fitness function, $f$, is then calculated as a measure of how well adapted the individual is.  The fitness function is essentially the inverse of the error, such that the individual's fitness is highest when it is closest to its flame and detonation property target values:

\begin{equation}
f = \frac{1}{{error + \varepsilon }},
\label{fitness}
\end{equation}

\noindent
where $\varepsilon$ is a very small number used to prevent division by zero.

New generations are then created, in which there is a 60\% probability of crossover and a 10\% probability of mutation.  Crossover occurs when traits from two fit  individuals are transferred to the next generation, forming an even more highly adapted individual.  Mutation occurs when random changes in certain traits are passed to the next generation, resulting in either stronger or weaker individuals.  In this way, the characteristics of those individuals with high fitness functions are passed on to the next generation, while weak individuals are eliminated.  

The procedure described above is applied to each of the subdivisions for a user-specified number of generations.   Then, as indicated in the third block of Fig.\ref{overall_ga}, all of the individuals from each subdivision are combined into one main population, and the procedure is repeated for all of the individuals in the main population for a user-specified number of generations.  The individual with the highest fitness function is then chosen as the best set of reaction parameters from the GA.  

A genetic algorithm is a good optimization tool because it is able to search over a wide range of parameter combinations for a best solution, and because it does not depend on initial estimates, which prevents the solution process from being trapped in local optima. This independence arises partly because a genetic algorithm uses random-number generation when calculating the values of reaction parameters in the initial population and in the crossover and mutation probabilities.  The optimal traits of an individual however, may be in a narrow region, which is missed due to this random number generation.    Because of this, {\sl{the GA might find solutions close to an optimum, rather than the optimum itself.}}  Therefore, to improve the optimization procedure, the result from the GA is used as input into a Nelder-Mead (NM) optimization scheme.

\subsection{Application of a Nelder-Mead Optimization Scheme}

The NM algorithm \cite{lagarias_convergence_1998} is an optimization scheme developed for multivariable minimization problems. Unlike the GA, the NM procedure {\sl{is}} sensitive to the initial estimate, such that it can find a local optimum, rather than a global one.  Therefore, NM is used here to find the optimal solution after GA has provided a sufficiently close solution.

\begin{figure}[!htbp]
		\centering
		\includegraphics[width=1.0\textwidth]{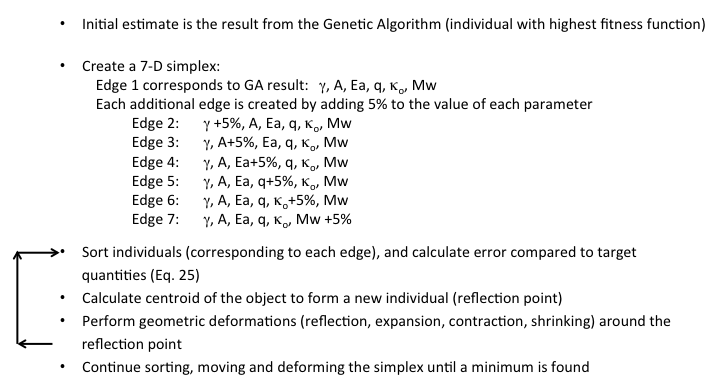}
		\centering
		\caption{Overall procedure for the Nelder-Mead optimization scheme}
		\label{overall_nm}
\end{figure}

Figure \ref{overall_nm} shows the NM procedure.  First, a simplex is created around an initial estimate. This simplex is a geometric structure consisting of n+1 edges, where n represents the number of variables.  In this case, there are six reaction parameters ($\gamma$, $A$, $E_a$, $q$, $\kappa_o$, and $M_w$), and so there are seven edges.  The first edge is defined by the initial estimate, which is the genetic algorithm result. Each of the other edges represents an individual created by adding 5\% to the value of one of the variables, while keeping the others the same. For example, edge 2 is formed by increasing the value of $\gamma$ from the initial estimate by 5\% and keeping the values for the other other variables same, while edge 3 is formed by increasing value of $A$ by 5\%, while keeping the values for the rest of the parameters, including $\gamma$, the same as in the initial estimate. Once the 7-D simplex is created, these individuals are sorted based on their error, defined by Eq.\ref{error}.  A new individual is formed by averaging the values, while neglecting the value with highest error.  This individual is used to determine a point called the ``reflection point.''   Based on the error of this reflection point, compared with the initial estimate, the following actions are taken: reflection, expansion, contraction, inside contraction, or shrinking around this point.  With these steps, the algorithm moves and deforms the simplex to find a minimum. 

\section{Results}

\subsection{Procedure Setup}
\label{procedure_setup}

Here, we use the GA-NM procedure to calculate optimal reaction parameters appropriate for simulations of flame acceleration and DDT in stoichiometric methane-air and ethylene-oxygen mixtures.   For both mixtures, the initial ambient conditions are $P_o$ = 1 atm, and $T_o$ = 298 K.  As discussed below, prespecified target flame and detonation properties are calculated and are shown in Table \ref{table_target_values}.

Flame temperatures, $T_b$ and $T_{cv}$ are determined from Cantera \cite{Cantera} chemical equilibrium software, using the GRI 3.0 \cite{smith_gri-mech} mechanism, and validated with the Gordon McBride \cite{Gordon_McBride} equilibrium program.  Laminar flame speed and thickness, $S_l$ and $x_{ft}$, are calculated from Cantera's \cite{Cantera} freely-propagating premixed flame software, where $x_{ft}$ is calculated from the 1-D flame profile, using Eq. \ref{xf}.   The Chapman\--Jouguet detonation velocity $D_{CJ}$ is calculated using the Shock and Detonation Toolbox software  \cite{california_institute_of_technology_shock_2015}, using the GRI 3.0 \cite{smith_gri-mech} mechanism, and validated with the Gordon McBride \cite{Gordon_McBride} equilibrium program.  The half reaction thickness, $x_d$ is set equal to the distance to peak thermicity calculated from the ZND profile \cite{california_institute_of_technology_shock_2015}.

\begin{table}
\tbl{Pre-specified target flame and detonation properties for stoichiometric methane-air and ethylene-oxygen mixtures.}
{\begin{tabular}{l c c}  \toprule 
Target Property & {Methane-Air} & {Ethylene-Oxygen} \\
\midrule
 {$T_b$} (K) & 2224 & 3173 \\     
    {$S_l$} (cm/s) & 37.7 & 482.2   \\     
    {$x_{ft}$} (cm) & 4.46 x $10^{-2}$ & 6.51 x $10^{-3}$   \\     
    {$T_{cv}$} (K) & 2585 & 3731   \\     
    {$D_{CJ}$} (m/s) & 1800 & 2373   \\       
     {$x_d$} (cm) & 1.68 & 2.15 x $10^{-3}$  \\     
\bottomrule
    \end{tabular}}
\label{table_target_values}
\end{table}

Following the procedure discussed in Section \ref{optimization_procedure}, an initial population is created.  The upper and lower bounds and number of subdivisions for each parameter have been shown in Table \ref{table_bounds}.  Forty individuals are created within each subdivision, and each subdivision undergoes its own GA procedure for 100 generations.  Then all of the individuals from all subdivisions are combined into one main population, which undergoes a GA procedure for 1000 generations.  During the GA procedure, as each new generation is formed, the probability of crossover and mutation are 60\% and 10\%, respectively.  As discussed in Section \ref{optimization_procedure}, the result from the genetic algorithm is used as the starting point for the Nelder-Mead procedure.   

\subsection{Optimal Reaction Parameters for Stoichiometric Methane-Air and Ethylene-Oxygen Mixtures}
\label{optimal_reaction_params_in_results}

The results from the GA-NM procedure for stoichiometric methane-air and ethylene-oxygen mixtures are shown in Table \ref{table_reaction_params}.  These are the optimal reaction parameters that reproduce the prespecified target flame and detonation properties listed in Table \ref{table_target_values}.   Table \ref{table_comparison} compares the target flame and detonation properties and the value of these properties when the reaction parameters are used in 1-D flame and detonation fluid dynamics codes.  The error, as described by Eq. \ref{error}, is 0.61\% for stoichiometric methane-air and 1.27\% for stoichiometric ethylene-oxygen.  This indicates that the reaction parameters computed from GA-NM, when used in Navier-Stokes simulations, successfully reproduce the target flame and detonation properties.  

The amount of CPU time required for the GA-NM procedure depends heavily on the number of subdivisions of each reaction parameter, the number of individuals in each subdivision, and the number of generations computed within each subdivision and for the entire main population.  For the GA-NM calculations shown here, the wall-clock time is approximately four hours on a single core of a Dell mini-cluster.

\begin{table}
\tbl{Optimal reaction parameters from GA-NM procedure for stoichiometric methane-air and ethylene-oxygen mixtures.}
{\begin{tabular}{l c c} \toprule 
Optimal Reaction & Methane-Air & Ethylene-Oxygen \\
Parameters & & \\   \midrule
      {$\gamma$}  & 1.18   &  1.19 \\ 
    {$A$} (cm$^3$/g$\cdot$s) & {8.76 x ${10^{13}}$}  &   {2.16 x ${10^{15}}$}  \\ 
   {$E_a$} & {$79.17RT_o$}  & {$92.80RT_o$}  \\
    {$q$} & {$42.29RT_o/{M_w}$} &  {$57.85RT_o/{M_w}$}  \\  
    {$\kappa_0$} (g/s$\cdot$cm$\cdot$K$^{0.7}$) & {6.89 x ${10^{-6}}$} & {8.63 x ${10^{-6}}$}  \\     
    {$M_w$} (g/mole) & {27.29}   & {23.67}  \\  
    \bottomrule
    \end{tabular}}
\label{table_reaction_params}
\end{table}

\begin{table}
\tbl{Comparison between target properties (from Cantera \cite{Cantera} and Shock \& Detonation Toolbox \cite{california_institute_of_technology_shock_2015}) and properties obtained when using the optimal reaction parameters in a 1-D fluid code.}
{\begin{tabular}{l c c c c}
\toprule 
 & \multicolumn{2}{c}{{Methane-Air}} & \multicolumn{2}{c}{{Ethylene-Oxygen}} \\
                \cmidrule(lr) {2-3} \cmidrule(lr){4-5}

               & {{Target}} & {{Property when}}  &  {{Target }} & {{Property when}} \\ 
             & {{properties }} & {{using optimal}}  & {{properties}} & {{using optimal}}  \\               
            & {{from}} & {{reaction parameter}}  &  {{from}} & {{reaction parameter}} \\ 
          & {{Table \ref{table_target_values}}} & {{in 1-D fluid code}}  & {{Table \ref{table_target_values}}} & {{in 1-D fluid code}}  \\  
               \cmidrule(lr) {2-3} \cmidrule(lr){4-5}

       {$T_b$} (K) & 2224 & 2229 & 3173 & 3167  \\     
    {$S_l$} (cm/s) & 37.7 & 37.7 & 482.2 &  483.5  \\     
    {$x_{ft}$} (cm) & 4.46 x $10^{-2}$ & 4.48 x $10^{-2}$ & 6.51 x $10^{-3}$ &  6.47 x $10^{-3}$   \\     
    {$T_{cv}$} (K) & 2585 & 2579 &  3731 & 3739   \\     
    {$D_{CJ}$} (m/s) & 1800 &1800 & 2373 & 2359   \\       
     {$x_d$} (cm) & 1.68 & 1.67 & 2.15 x $10^{-3}$ & 2.17 x $10^{-3}$  \\   
     
                    \cmidrule(lr) {2-3} \cmidrule(lr){4-5} 
 & \multicolumn{2}{c}{{Error = 0.61\%}} & \multicolumn{2}{c}{{Error = 1.27\%}}  \\ 
    \bottomrule
    \end{tabular}}
\label{table_comparison}
\end{table}

\section{Discussion}

\subsection{Evaluation of Optimal Reaction Parameters in a Fluid-Dynamics Code}

The GA-NM procedure finds six optimal reaction parameters ($\gamma$, $A$, $E_a$, $q$, $\kappa_o$, and $M_w$) that reproduce six target flame and detonation properties ($T_b$, $S_l$, $x_{ft}$, $T_{cv}$, $D_{CJ}$, and $x_d$).   Because these target parameters include {\sl{both}} flame and detonation properties, these optimal reaction parameters {\sl{should theoretically}} reproduce the major characteristics of the system when used in a fluid code to simulate flame acceleration and DDT.  As shown below, however, flame acceleration and DDT are complex processes, and, in this section, we examine if one set of optimal reaction parameters is consistent or correct for the whole evolution of the flow.  

\begin{figure}
		\centering
		\includegraphics[width=1.0\textwidth]{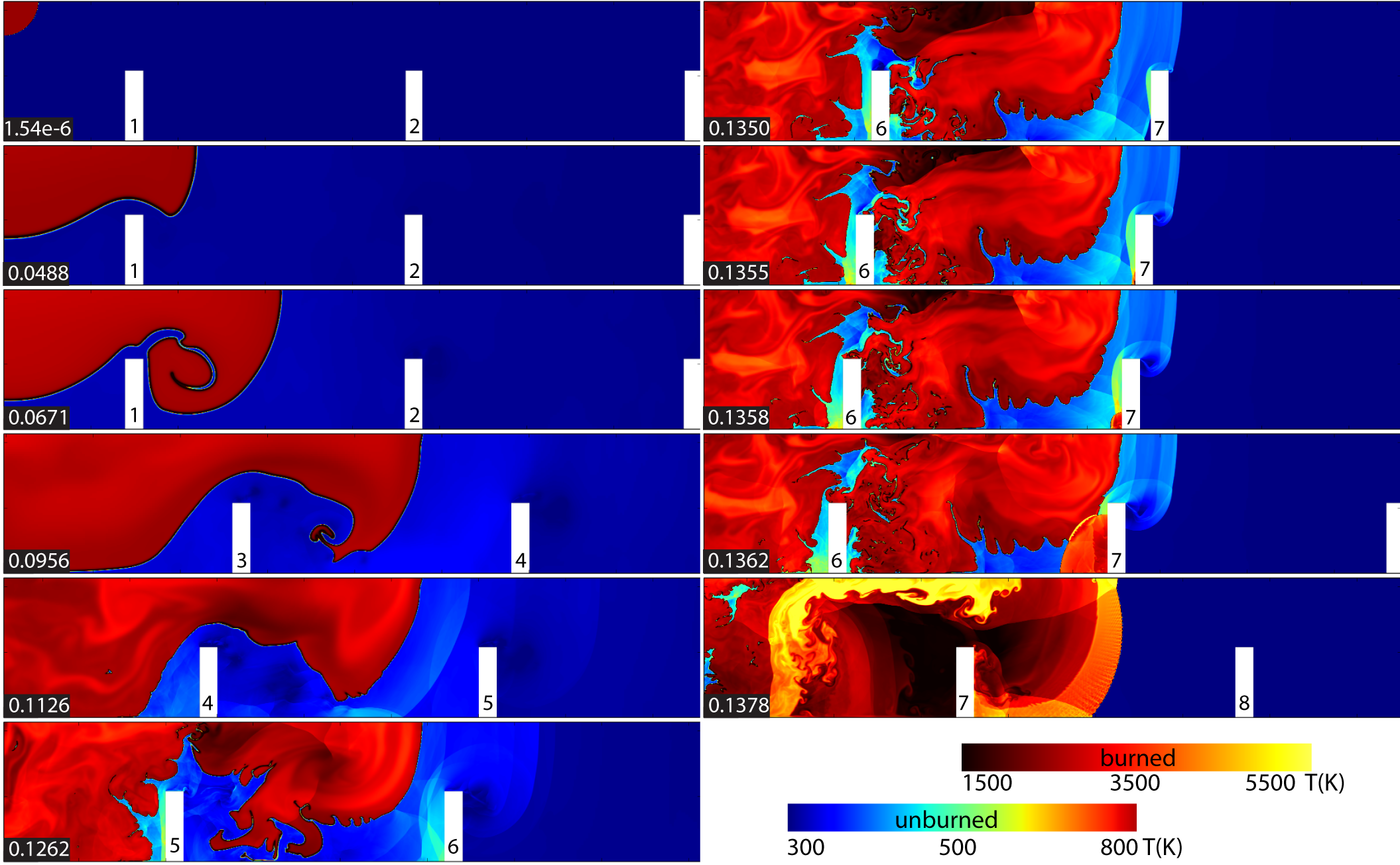}
		\centering
		\caption{Sequence of temperature images from a simulation of flame acceleration and DDT for a stoichiometric mixture of ethylene-air, in a channel with a blockage ratio of 0.5.  Figure courtesy of Gabriel Goodwin \cite{goodwin_2016, goodwin_2017}.}
		\label{gabe_DDT}
\end{figure}

Goodwin et al. \cite{goodwin_2016, goodwin_2017} have simulated DDT in long rectangular channels containing a mixture of stoichiometric ethylene-air.  The channel contained regularly spaced obstacles and the simulations were used to quantify the effects of decreasing the blockage ratio of those obstacles on the DDT mechanism \cite{goodwin_2016}.  Details of the computational geometry and simulation conditions can be found in \cite{goodwin_2016,goodwin_2017}.  Figure \ref{gabe_DDT} shows a sequence of temperature images from a simulation in which the obstacle blockage ratio is 0.5.  The images are labeled by physical time, in {\sl{ms}}, in the lower left hand corner of each frame, and the  obstacles are labeled by number.  The computational domain has a symmetry boundary condition at the top and a wall boundary condition along the bottom of each frame, and the obstacle surfaces are walls.  As shown in the first frame, {\sl{1.54 x $10^{-6}$ ms}}, the flame is initiated at the upper left boundary.  Subsequent frames show that the flame accelerates as it passes over obstacles, and its surface area increases due to expansion and fluid instabilities (Rayleigh\--Taylor, Richtmyer\--Meshkov, and Kelvin\--Helmholtz).  At {\sl{0.1126 ms}}, a shock forms ahead of the flame, and by {\sl{0.1262 ms}}, the flame has become increasingly turbulent due to the fluid instabilities.  The strong leading shock propagates into the unburned gas, diffracts over obstacles and reflects from the channel floor, eventually forming a Mach stem at {\sl{0.1350  ms}}, in front of obstacle 7.  At  {\sl{0.1355 ms}}, the Mach stem reflects from obstacle 7, causing a pressure and temperature increase behind the Mach reflection, which then leaves a hot spot in a gradient of reactivity.  Finally, the detonation occurs in the hot, unburned gas behind the Mach stem reflection.

\begin{figure}
		\centering
		\includegraphics[width=0.7\textwidth]{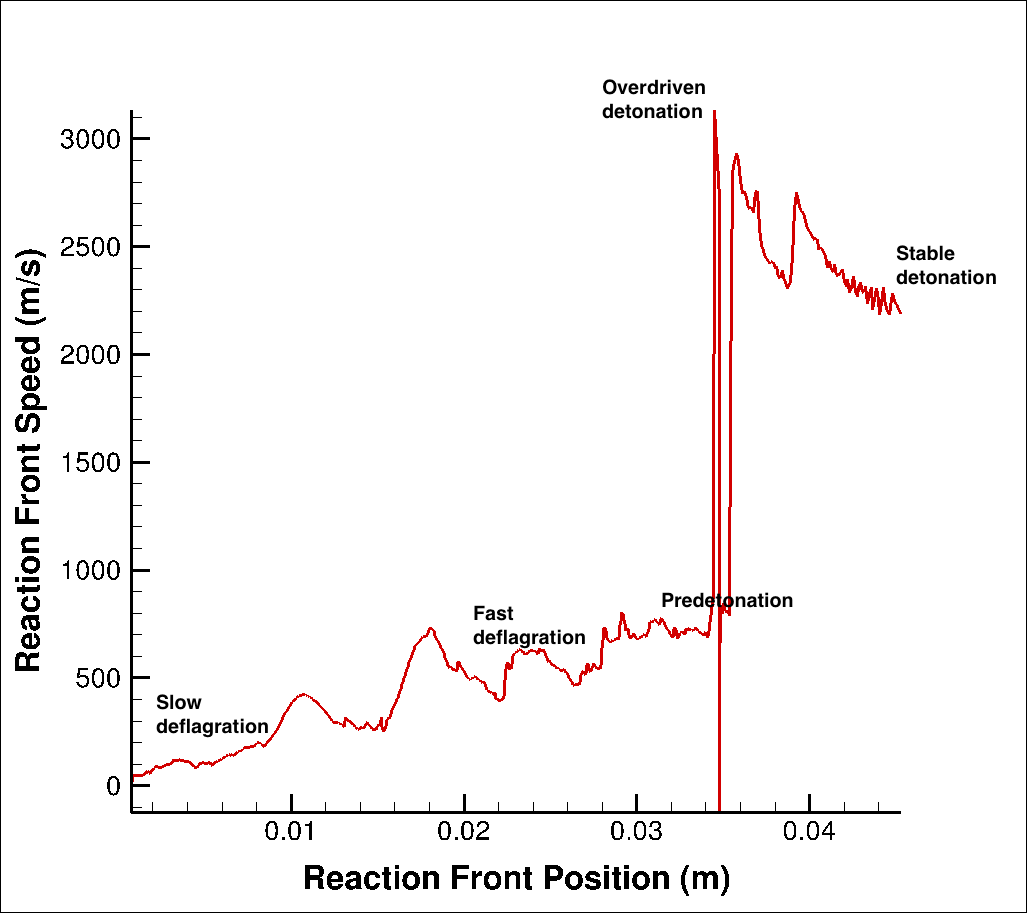}
		\centering
		\caption{Reaction front position vs. speed for the image sequence shown in Fig. \ref{gabe_DDT}.  Figure courtesy of Gabriel Goodwin \cite{goodwin_2016, goodwin_2017}.}
		\label{position_vs_speed}
\end{figure}

The process of DDT can be separated into five phases, as listed below, and as annotated in Fig. \ref{position_vs_speed}:

\begin{itemize}
\item
{{\sl{Slow deflagration phase}}, in which the flame speed is subsonic and increases approximately to the speed of sound in the mixture. The increase of deflagration speed is primarily due to the development of background turbulence and its interaction with the deflagration front.  This corresponds to the first three frames in Fig \ref{gabe_DDT} in which $0<t<0.0671$  ms and at the earliest segment of the line plot in Fig \ref{position_vs_speed}.}
\smallskip
\item
{{\sl{Fast deflagration phase}}, also called the Òfast flameÓ stage, and which can be seen in Fig \ref{gabe_DDT} at {\sl{0.0956} ms}. Here the accelerating turbulent flame is supersonic and begins to generate shock waves. The shock waves compress the background flow ahead of the deflagration, and decrease the reaction time. The result is that the turbulent deflagration speed increases further, to the point where it is ~0.5$D_{CJ}$ of the fuel-air mixture. }
\smallskip
\item
{{\sl{Predetonation phase}}, in which shocks and shock interactions create conditions in which a detonation might occur.  This can be see in Fig \ref{gabe_DDT} at $0.1126<t<0.1355$  ms and immediately before the sharp spike in Fig \ref{position_vs_speed}.}
\smallskip
\item
{{\sl{Overdriven detonation}}, which occurs when conditions behind a leading shock front reach a critical value and the system transitions to a detonation. Immediately after this transition, the detonation is originally overdriven and the local pressure at the DDT site can become very large, as shown in $0.1358<t<0.1362$  ms and at the peak in Fig \ref{position_vs_speed}}.
\smallskip
\item
{{\sl{Stable detonation propagation}}, in which the steady detonation propagates at a speed of around $D_{CJ}$, until it runs out of fuel or is disrupted in some way, as shown at {\sl{0.1378} ms} and the last annotation in Fig \ref{position_vs_speed}.}
\end{itemize}

As shown above, flame acceleration and DDT involves thermal expansion, turbulence, fluid instabilities, shock-flame interactions, shock collisions and reflections.  The GA-NM procedure produces one set of reaction parameters, ($\gamma$, $A$, $E_a$, $q$, $\kappa_o$, and $M_w$) to simulate all five phases of flame acceleration and DDT.  As indicated in Table \ref{table_reaction_params}, for ethylene-oxygen mixtures, the overall optimal $\gamma$ is 1.19, and the overall $M_w$ is 23.67 g/mole.  In actuality, quantities such as $\gamma$ and $M_w$ vary throughout the entire computational domain and throughout the different phases of DDT.  One-dimensional flame and detonation programs using detailed chemical mechanisms produce different values for $\gamma$ and $M_w$ for the burned and unburned regions.  For example, Gordon-McBride \cite{Gordon_McBride} and Shock \& Detonation Toolbox \cite{california_institute_of_technology_shock_2015} detonation calculations using the GRI mechanism \cite{smith_gri-mech} predict that $\gamma$ in the unburned and burned gases are 1.34 and 1.14, respectively.   The Cantera \cite{Cantera} constant-enthalpy and constant-pressure equilibrium calculations using the GRI mechanism predict that $M_w$ for reactants and products are 31.01 g/mole and 23.17 g/mole, respectively.  This indicates that the single, overall  $\gamma$ and $M_w$ predicted by GA-NM ($\gamma$=1.19 and $M_w$=23.67 g/mole) are shifted towards the products, rather than the reactants.  Therefore, it may be that different values for overall $\gamma$ and $M_w$ would be appropriate for the different phases of flame acceleration and DDT shown in Figs. \ref{gabe_DDT} and \ref{position_vs_speed}, and that the GA-NM model could be tuned to calculate optimal reaction parameters for different regions of the flow.

\subsection{Comparison between GA-NM, Graphical Approach and Experimental Measurements for Stoichiometric Methane-Air Mixtures}
\label{compare_GA-NM_graphical}

\begin{table}
\tbl{Comparison of reaction parameters from the Graphical Approach \cite{kessler_simulations_2010} and from GA-NM for stoichiometric methane-air.}   
{\begin{tabular}{l c c}  \toprule 
{Reaction Parameter} & {Graphical Approach} & {GA-NM} \\
  \midrule
 $\gamma$ & 1.197 &  1.181   \\     
 {$A$} (cm$^3$/g$\cdot$s) &1.64 x $10^{13}$ &  8.76 x $10^{13}$   \\  
{$E_a$} & {$67.55RT_o$} &  {$79.17RT_o$}   \\         
 {$q$} & {$39.0RT_o/{M_w}$}  & {$42.3RT_o/{M_w}$}   \\  
  {$\kappa_0$} (g/s$\cdot$cm$\cdot$K$^{0.7}$) & 6.25x $10^{-6}$  &  6.89 x $10^{-3}$  \\     
  {$M_w$} (g/mole) & 27.0  & 27.3  \\     
\bottomrule
\end{tabular}}
\label{Graphical_Approach_vs_GA-NM}
\end{table}

To evaluate the reaction parameters computed by their graphical approach, Kessler et al. \cite{kessler_simulations_2010} used their parameters in 2-D simulations of flame acceleration and DDT in obstacle-laden channels containing stoichiometric methane-air and compared their results with available experimental measurements \cite{johansen_2009,kuznetsov_2002}.  Simulations were conducted with various blockage ratios to match the experimental configurations \cite{johansen_2009, kuznetsov_2002}.   Their results \cite{kessler_simulations_2010} showing reaction front velocity vs.\ reaction front position for a case in which the channel has a diameter of 17.4 cm and a blockage ratio {\sl{br = 0.3}} \cite{kuznetsov_2002}, indicate that their reaction parameters reproduce the experimental observations, both qualitatively and quantitatively.  They \cite{kessler_simulations_2010} showed that varying the model parameters to produce relatively small (10-15\%) changes in individual laminar flame properties had little impact on the observed flame acceleration and DDT and surmised that any differences were due to chance fluctuations in the thermodynamic conditions within the hot spots that initiate detonations.  

Table \ref{Graphical_Approach_vs_GA-NM} shows a comparison of the reaction parameters found by the graphical approach \cite{kessler_simulations_2010} and those found in this study using the GA-NM procedure.  The most significant differences are in the values of the preexponential factor, $A$ and and activation energy $E_a$.  Although the GA-NM activation energy is higher (which would lower the reaction rate), the preexponential factor is also higher (which would increase the reaction rate), and therefore the differences in these two parameters have competing effects.  It is important to note that multiple sets of reaction parameters can reproduce the flame and detonation target values, and that the reaction parameters produced by the GA-NM procedure here correspond to the set of parameters that maximize the genetic algorithm's fitness function, as specified in Eq. \ref{fitness}.

The differences in the reaction parameters may also be attributed to differences in some of the values of the target parameters used by the two methods.  As discussed in Sec. \ref{procedure_setup}, the target parameters used in this study, which are shown in Table \ref{table_target_values}, were obtained from equilibrium calculations, and 1-D flame and detonation calculations \cite{Gordon_McBride, Cantera, california_institute_of_technology_shock_2015}.   The target parameters used in the graphical approach \cite{kessler_simulations_2010} were based primarily on experiments, and are: $T_b$=2210K, $S_l$=38.02 cm/s, $D_{CJ}$=1820 m/s, and $x_d$=0.229 cm.   Hence, the target values used in the graphical approach for $T_b$, $S_l$, and $D_{CJ}$ are very similar to those used here in the GA-NM procedure, but the target value used for $x_d$ is very different.  (In the graphical procedure \cite{kessler_simulations_2010}, experimental measurements of detonation cell size, $\lambda$ were used \cite{kuznetsov_2002}, and then $x_d$ was estimated based on the relationship that $50<\lambda /{x_d}  <100$; whereas, in this study using the GA-NM procedure, the half reaction thickness, $x_d$ is set equal to the distance to peak thermicity calculated from the ZND profile \cite{california_institute_of_technology_shock_2015}.)

\begin{figure}[!htbp]
		\centering
		\includegraphics[width=0.8\textwidth]{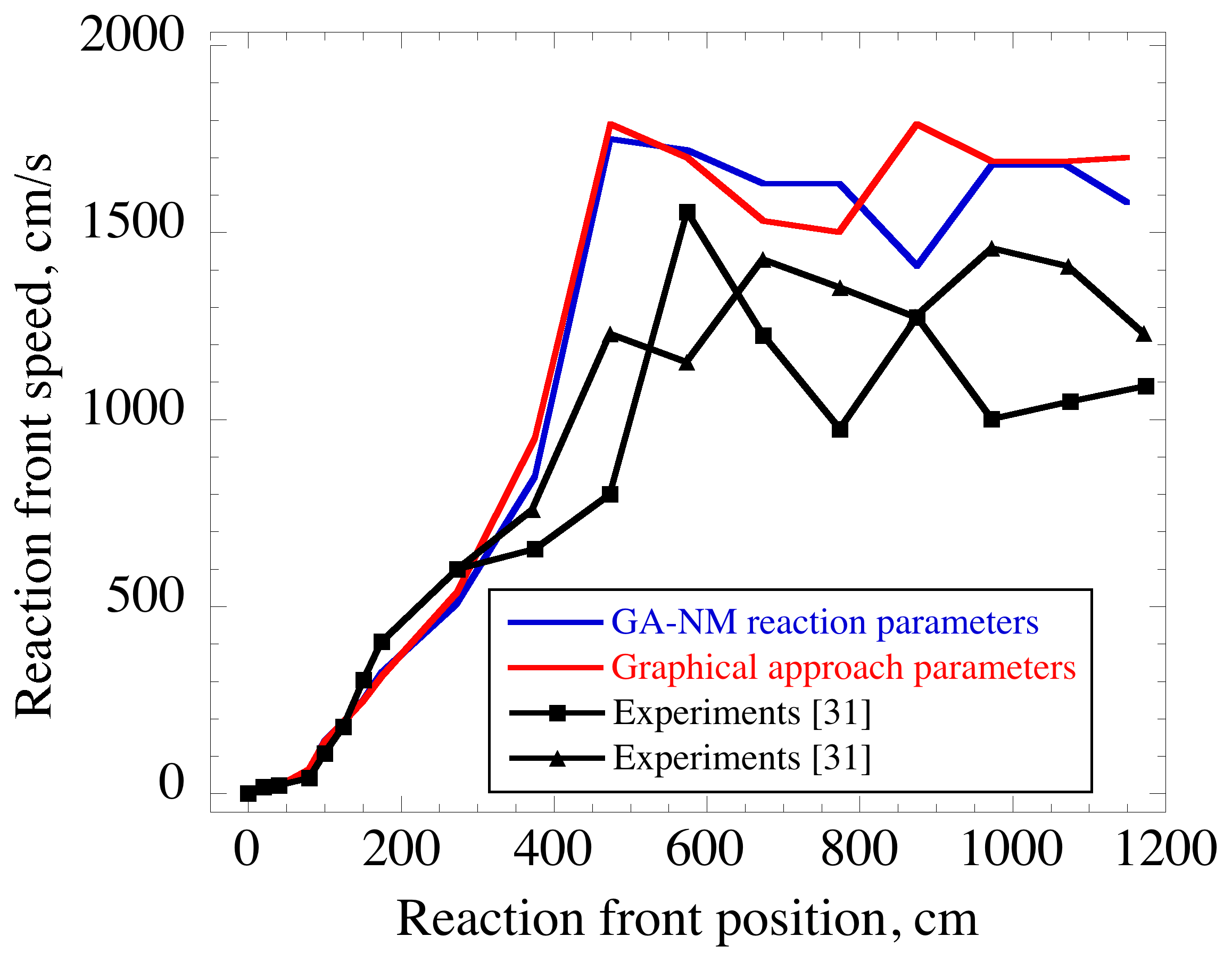}
		\centering
		\caption{Reaction front position vs.\ reaction front speed from a 2-D simulation of flame acceleration and DDT }
		\label{kuznetsov_comparison}
\end{figure}

The reaction parameters from the GA-NM procedure and from the graphical approach were used as inputs in 2-D simulations of flame acceleration and DDT in an obstacle-laden channel, containing stoichiometric methane-air.  The channel had a diameter of 17.4 cm and blockage ratio {\sl{br = 0.3}} to match the experimental configuration of Kuznetsov \cite{kuznetsov_2002}.  Figure \ref{kuznetsov_comparison} shows reaction front position vs.\ reaction front speed for these simulations using FAST (Flame Acceleration Simulation Tool) \cite{houim_2016}, in which the mesh was dynamically refined around shocks, flame fronts and in regions of large gradients of density, pressure or composition.  In addition to the simulation results, Fig. \ref{kuznetsov_comparison} also shows reaction front location vs.\ reaction front speed data from the experiments \cite{kuznetsov_2002}.  Despite the differences in the GA-NM vs.\  graphical approach reaction parameters, the simulations using parameters from both methods closely agree with each other, and both follow the transition to detonation observed in the experiments.  This is most likely because both sets of reaction parameters reproduce the target flame and detonation properties, even though some the parameters are different.

\subsection{Uncertainties in Target Values}

The reaction parameters shown in Table \ref{table_reaction_params} are strongly affected by the target parameters on which they are based, shown in Table \ref{table_target_values}.  That is, the user provides values for the target parameters (these are inputs to the GA-NM algorithm), based on reasonable or best numbers available, such as from experiments or simulations using detailed chemical reaction mechanisms.  The GA-NM output is the set of reaction parameters that most closely match the target parameters and have the smallest formal error, based on Eq. \ref{error}.  

As discussed in Section \ref{procedure_setup}, the prespecified target flame and detonation properties are calculated using the Gordon McBride chemical equilibrium software \cite{Gordon_McBride}, Cantera \cite{Cantera}, and the Shock and Detonation Toolbox \cite{california_institute_of_technology_shock_2015}, using the GRI 3.0 \cite{smith_gri-mech} mechanism.  Changing the target values has a significant effect on the resulting optimal reaction parameters calculated by the GA-NM algorithm, because the fitness of a set of reaction parameters is based on how closely matched the target values are to the prespecified values.  To the extent that different chemical reaction mechanisms produce different values for some flame and detonation properties, the choice of which mechanism to use to calculate target properties can affect the resulting GA-NM reaction parameters.

Most of the target quantities used in the GA-NM algorithm ($T_b$, $S_l$, $x_{ft}$, $T_{cv}$, and $D_{CJ}$) are fairly well established quantities for hydrocarbon-air mixtures.  The quantity with the most uncertainty is the half-reaction thickness, $x_d$.  As discussed in Sections \ref{procedure_setup} and \ref{compare_GA-NM_graphical}, for the GA-NM calculations presented in this paper, the value for $x_d$ is taken to be the distance to peak thermicity, as calculated in the ZND profile from the Shock and Detonation Toolbox \cite{california_institute_of_technology_shock_2015} program.  Other studies \cite{kessler_simulations_2010,alp_scholarly_paper} have used experimental measurements of detonation cell size, $\lambda$, and then estimated $x_d$.  For target quantities with large uncertainties, it is beneficial to minimize the effect of that lesser-known quantity on the error (Eq. \ref{error}) that is used to calculate the fitness function (Eq. \ref{fitness}).  In GA-NM, this is done by incorporating a weighting factor, $wf_i$:

\begin{equation}
\label{weighted_error}
{Error}=\left[ \sum_{i=1}^{ntargets} \left(wf_i \cdot  \frac {\xi_i-\xi_{i,target}} {\xi_{i,target}} \right)^2 \right]^{1/2}.
\end{equation}

\noindent
This weighting factor can be set to a small value (less than unity) for quantities with larger errors, which will reduce the impact of the uncertainty in the target parameter on the GA-NM optimal reaction parameters.  

Although the six target quantities used in this study were $T_b$, $S_l$, $x_{ft}$, $T_{cv}$, $D_{CJ}$, and $x_d$, any other quantity can be used as a target property.  As part of the GA-NM procedure during which flame and detonation properties are calculated (corresponding to  Eqs.~\ref{flame1}-\ref{flame3} and Eqs.~\ref{zndstart}-\ref{zndend}), the algorithm tracks other properties such as CJ quantities $T_{CJ}$, $P_{CJ}$, $\rho_{CJ}$, and post-shock quantities $T_{ZND}$, $P_{ZND}$, $\rho_{ZND}$.  These could be used as target quantities, either in place of or in addition to the six used here.  Another useful quantity to use as a target property is detonation cell size, $\lambda$, which could be obtained from experimental measurements \cite{kuznetsov_2002}.  Using $\lambda$ as a target property would ensure that correct detonation cell sizes would be obtained when the corresponding optimal reaction parameters are used in Navier-Stokes solvers.  

\section{Conclusions and Future Work}

One major problem in simulating flame acceleration and DDT is finding adequate models of the chemical reactions, heat release, and physical diffusion processes that are required source terms in the Navier-Stokes equations.  Currently available detailed chemical reaction mechanisms are both computationally prohibitive and inaccurate for high-temperature and high-pressure shock-laden flows.  This paper has presented an automated procedure to determine the reaction parameters for a chemical-diffusive model to simulate flame acceleration and DDT in stoichiometric methane-air and ethylene-oxygen mixtures.  

The new procedure uses a combination of a genetic algorithm and Nelder-Mead optimization scheme (GA-NM)  to find the optimal reaction parameters for a reaction rate based on a simplified Arrhenius type form of conversion of reactants to products.  The model finds six optimal reaction parameters, $\gamma$, $A$, $E_a$, $q$, $\kappa_o$, and $M_w$ that reproduce six target flame and detonation properties, $T_b$, $S_l$, $x_{ft}$, $T_{cv}$, $D_{CJ}$, and $x_d$.  Values for the target flame and detonation properties for methane-air and ethylene-oxygen mixtures were obtained from chemical equilibrium software and from 1-D flame and detonation calculations using detailed chemical reaction mechanisms.  

Results from this study show that the optimal reaction parameters, when used to solve the 1-D reactive Navier-Stokes equations, closely reproduce those target flame and detonation properties for both mixtures.  The effects of uncertainties in the values of target flame and detonation properties can be minimized to have little effect on the resulting optimal reaction parameters, and the reaction parameters can be tailored, if necessary, for the different regimes of flame acceleration and DDT.  The GA-NM optimal reaction parameters, when used as input in a 2-D simulation of flame acceleration and DDT in an obstacle-laden channel containing stoichiometric methane-air, closely follow the transition\--to\--detonation observed in experiments \cite{kuznetsov_2002}.  This automated procedure to determine reaction parameters for any fuel-air mixture makes it possible to simulate flame acceleration and DDT in large channels, calculations that would otherwise be incalculable.

Near-term future work includes evaluation of using alternative properties for target parameters, such as detonation cell size, $\lambda$.  Additional future work includes using the GA-NM method to find optimal reaction parameters for fuel-air mixtures with spatially varying equivalence ratios.  This would be applicable to simulate scenarios of flame acceleration and DDT in coal mines containing regions with high- and low concentrations of methane.  In addition, future studies will use the GA-NM procedure to find reaction parameters for other fuels, and more complex (possibly multistep) forms for representing the evolution of the energy release.

\section*{Acknowledgments}

The authors are grateful to Gabriel Goodwin for providing Figs. \ref{gabe_DDT} and  \ref{position_vs_speed}, to Drs.\ Ryan Houim, Weilin Zheng, and Huahua Xiao for Navier-Stokes (FAST) simulations using the GA-NM reaction parameters, and to Drs.\ David Kessler and Vadim Gamezo for their background work in development of chemical-diffusive models.

\section*{Disclosure statement}
The authors have no potential conflicts of interest.

\section*{Funding}
This work was supported in part by the University of Maryland through Minta Martin Endowment Funds in the Department of Aerospace Engineering, and through the Glenn L. Martin Institute Chaired Professorship at the A. James Clark School of Engineering.  In addition, this work was supported by the Alpha Foundation (Grant No. AFC215--20) and the Office of Naval Research (Contract No. N00014--14--1--0177).  The authors acknowledge the University of Maryland supercomputing resources (http://www.it.umd.edu/hpcc) made available in conducting the research reported in this paper.    




\begin{thebibliography}{99}

\bibitem{oran_origins_2007}
E.S. Oran and V.N. Gamezo, \emph{Origins of the deflagration-to-detonation transition in gas-phase combustion}, Combustion and Flame 148 (2007), pp. 4-47.

\bibitem{zipf_2013}
R.K. Zipf, V.N. Gamezo, M.J. Sapko, W.P. Marchewka, K.M. Mohamed, E.S. Oran, D.A. Kessler, E.S. Weiss, J.D. Addis, F.A. Karnack, and D.D. Sellers, \emph{Methane-air detonation experiments at NIOSH Lake Lynn laboratory}, Journal of Loss Prevention in the Process Industries 26 (2013), pp. 295-301.

\bibitem{zipf_report}Zipf Jr, RK, Sapko, MJ, and Brune, JF, \emph{Explosion pressure design criteria for new seals in U.S. coal mines}, Pittsburgh, PA, U.S. Dept. HHS, NIOSH IC 9500 Report, 2007.

\bibitem{ciccarelli_2008}G. Ciccarelli and S. Dorofeev, \emph{Flame acceleration and transition to detonation in ducts}, Progress in Energy and Combustion Science 34 (2008), pp. 499-550.

\bibitem{roy_pulse_2004}
G.D. Roy, S.M. Frolov, A.A. Borisov, and D.W. Netzer, \emph{Pulse detonation propulsion: challenges, current status, and future perspective}, Progress in Energy and Combustion Science 30 (2004), pp. 545-672.

\bibitem{ju_microscale_2015}
Y. Ju, C.P. Cadou, and K. Maruta, \emph{Microscale Combustion and Power Generation}, Engineering collection; Engineering collection., Momentum Press, New York, [New York] (222 East 46th Street, New York, NY 10017), 2015.

\bibitem{nordeen_rde}
C.A. Nordeen, D. Schwer, F. Schauer, J. Hoke, T. Barber, and B. Cetegen, \emph{Thermodynamic model of a rotating detonation engine}, Combustion, Explosion, and Shock Waves 50 (2014), pp. 568-577.

\bibitem{wang_comprehensive_1998}
H. Wang and A. Laskin, \emph{A comprehensive kinetic model of ethylene and acetylene oxidation at high temperatures}, Internal Report, 1998.

\bibitem{kazakov_1995}
A. Kazakov, H. Wang, and M. Freklach, \emph{Detailed modeling of soot formation in laminar premixed ethylene flames at a pressure of 10 bar}, Combustion and Flame 100 (1995), pp. 111-120.

\bibitem{taylor_numerical_2013}
B.D. Taylor, D.A. Kessler, V.N. Gamezo, and E.S. Oran, \emph{Numerical simulations of hydrogen detonations with detailed chemical kinetics}, Proceedings of the Combustion Institute 34 (2013), pp. 2009-2016.

\bibitem{khokhlov_numerical_1999}
A.M. Khokhlov and E.S. Oran, \emph{Numerical simulation of detonation initiation in a flame brush: the role of hot spots}, Combustion and Flame 119 (1999), pp. 400-416.

\bibitem{gamezo_influence_2001}
V.N. Gamezo, A.M. Khokhlov, and E.S. Oran, \emph{The influence of shock bifurcations on shock-flame interactions and DDT}, Combustion and Flame 126 (2001), pp. 1810-1826.

\bibitem{gamezo_2007}
V.N. Gamezo, T. Ogawa, and E.S. Oran, \emph{Numerical simulations of flame propagation and DDT in obstructed channels filled with hydrogen-air mixture}, Proceedings of the Combustion Institute 31 (2007), pp. 2463-2471.

\bibitem{gamezo_2008}
V.N. Gamezo, T. Ogawa, and E.S. Oran, \emph{Flame acceleration and DDT in channels with obstacles: Effect of obstacle spacing}, Combustion and Flame 155 (2008), pp. 302-315.

\bibitem{gamezo_deflagration--detonation_2009}
V.N. Gamezo, T. Ogawa, and E.S. Oran, \emph{Deflagration-to-detonation transition in H2-air mixtures: Effect of blockage ratio}, AIAA Paper 440 (2009), Available at http://arc.aiaa.org/doi/pdf/10.2514/6.2009-440.

\bibitem{ogawa_2013a}
T. Ogawa, V.N. Gamezo, and E.S. Oran, \emph{Flame acceleration and transition to detonation in an array of square obstacles}, Journal of Loss Prevention in the Process Industries 26 (2013), pp. 355-362.

\bibitem{ogawa_2013b}
T. Ogawa, E.S. Oran, and V.N. Gamezo, \emph{Numerical study on flame acceleration and DDT in an inclined array of cylinders using an AMR technique}, Computers and Fluids 85 (2013), pp. 63-70.

\bibitem{kessler_simulations_2010}
D.A. Kessler, V.N. Gamezo, and E.S. Oran, \emph{Simulations of flame acceleration and deflagration-to-detonation transitions in methane-air systems}, Combustion and Flame 157 (2010), pp. 2063-2077.

\bibitem{kessler_2012}
D.A. Kessler, V.N. Gamezo, and E.S. Oran, \emph{Gas-phase detonation propagation in mixture composition gradients}, Philosophical Transactions: Mathematical, Physical and Engineering Sciences 370 (2012), pp. 567-596.

\bibitem{alp_scholarly_paper}
A. \"{O}zgen, \emph{Optimizing simplified one-step chemical-diffusive models for deflagration-to-detonation transition calculations}, MS Paper, University of Maryland, 2016.

\bibitem{glassman_2008}
I. Glassman, \emph{Combustion}, Academic Press, Burlington, MA, 2008.


\bibitem{zeldovich_1950}
Y. Zeldovich, \emph{On the theory of the propagation of detonation in gaseous systems}, NACA Tech.
Memo No. 1261, 1950.

\bibitem{vonneumann_1963}
J. von Neumann, \emph{Theory of Detonation Waves}, John von Neumann 1903-1957 Collected
Works, Pergamon Press, Oxford, England, 1963.

\bibitem{doring_1943}
W. Doring, \emph{On the detonation processes in gases}, Ann Phys 43 (1943), pp. 421-436.

\bibitem{goldberg_genetic_1989}
D.E. Goldberg, \emph{Genetic algorithms in search, optimization, and machine learning}, Addison-Wesley Publishing Company, Reading, Mass., 1989.

\bibitem{Gordon_McBride}
S. Gordon and B. McBride, \emph{Computer program for calculation of complex chemical equilibrium composition, rocket performance, incident and reflected shocks and Chapman-Jouguet detonation}, NASA SP-273, 1971.

\bibitem{Cantera}
D.G. Goodwin, H.K. Moffat, and R.L. Speth, \emph{Cantera: An object-oriented software toolkit for
chemical kinetics, thermodynamics, and transport processes}, http://www.cantera.org (2016),
version 2.1.

\bibitem{california_institute_of_technology_shock_2015}
Explosion Dynamics Laboratory, California Institute of Technology, \emph{Shock \& Detonation Toolbox for Cantera 2.1} (2015).

\bibitem{smith_gri-mech}
G.P. Smith, D.M. Golden, M. Frenklach, N.W. Moriarty, B. Eiteneer, M. Goldenberg, C.T.
Bowman, R.K. Hanson, S. Song, W.C. Gardiner Jr., V.V. Lissianski, and Z. Qin, \emph{GRI-MECH
3.0}, Available at http://www.me.berkeley.edu/gri mech/.


\bibitem{lagarias_convergence_1998}
J. Lagarias, J. Reeds, M. Wright, and P. Wright, \emph{Convergence Properties of the Nelder-Mead
Simplex Method in Low Dimensions}, SIAM J. Optim. 9 (1998), pp. 112-147.

\bibitem{kuznetsov_2002}
M. Kuznetsov, G. Ciccarelli, S. Dorofeev, V. Alekseev, Y. Yankin, and T. Kim, \emph{DDT in methane-air mixtures},
Shock Waves 12 (2002), pp. 215-220.

\bibitem{johansen_2009}
C. Johansen and G. Ciccarelli, \emph{Visualization of the unburned gas flow field ahead of an accelerating flame in an obstructed square channel}, Combustion and Flame 156 (2009), pp. 405-416.

\bibitem{goodwin_2016}
G. Goodwin, R. Houim, and E. Oran, \emph{Effects of decreasing blockage ratio on {DDT} in small channels with obstacles}, Combustion and Flame 173 (2016), pp. 16-26.

\bibitem{goodwin_2017} G. Goodwin, R. Houim, and E. Oran, \emph{Shock transition to detonation in channels with obstacles}, Proceedings of the Combustion Institute 36 (2017), pp. 2717-2724.

\bibitem{houim_2016} R. W. Houim, A. \"{O}zgen, and E. S. Oran, \emph{The role of spontaneous waves in the deflagration- to-detonation transition in submillimetre channels}, Combustion Theory and Modeling 20 (2016) 1068Ð1087.

\end{thebibliography}

\end{document}